\PassOptionsToPackage{unicode}{hyperref}
\PassOptionsToPackage{hyphens}{url}
\documentclass[11pt,a4paper]{article}
\usepackage{amsmath,amssymb}
\usepackage{lmodern}
\usepackage{iftex}
\ifPDFTeX
  \usepackage[T1]{fontenc}
  \usepackage[utf8]{inputenc}
  \usepackage{textcomp}
\else
  \usepackage{unicode-math}
  \defaultfontfeatures{Scale=MatchLowercase}
  \defaultfontfeatures[\rmfamily]{Ligatures=TeX,Scale=1}
\fi

\usepackage[margin=1in]{geometry}
\usepackage{setspace}
\usepackage{booktabs}
\usepackage{tabularx}
\usepackage{multirow}

\usepackage{graphicx}
\usepackage{float}
\usepackage{caption}
\IfFileExists{microtype.sty}{
  \usepackage[]{microtype}
  \UseMicrotypeSet[protrusion]{basicmath}
}{}

\usepackage{xcolor}
\definecolor{linkblue}{RGB}{0,0,139}
\IfFileExists{bookmark.sty}{\usepackage{bookmark}}{\usepackage{hyperref}}
\hypersetup{
  colorlinks=true,
  linkcolor=linkblue,
  citecolor=linkblue,
  urlcolor=linkblue,
  pdfcreator={LaTeX via pandoc}}

\usepackage{titlesec}
\titleformat{\section}{\large\bfseries\uppercase}{\thesection.}{0.5em}{}
\titleformat{\subsection}{\normalsize\bfseries\itshape}{\thesubsection}{0.5em}{}

\title{\textbf{Enhancing Automatic Speech Recognition Through Integrated Noise Detection Architecture}}
\author{Karamvir Singh\\
\small\textit{Independent Researcher}\\
\small Email: karamvirsingh628@gmail.com}
\date{}

\begin{document}

\maketitle

\begin{abstract}
This research presents a novel approach to enhancing automatic speech recognition systems by integrating noise detection capabilities directly into the recognition architecture. Building upon the wav2vec2 framework, the proposed method incorporates a dedicated noise identification module that operates concurrently with speech transcription. Experimental validation using publicly available speech and environmental audio datasets demonstrates substantial improvements in transcription quality and noise discrimination. The enhanced system achieves superior performance in word error rate, character error rate, and noise detection accuracy compared to conventional architectures. Results indicate that joint optimization of transcription and noise classification objectives yields more reliable speech recognition in challenging acoustic conditions.
\end{abstract}

\textbf{Index Terms---}speech recognition, noise robustness, deep learning, wav2vec2, audio classification

\section{INTRODUCTION}

Modern automatic speech recognition systems have achieved remarkable performance through deep learning architectures, particularly models based on self-supervised learning paradigms. However, real-world deployment scenarios frequently involve challenging acoustic environments where background disturbances significantly compromise recognition accuracy. When processing audio containing substantial non-speech content, conventional systems often generate incoherent outputs, leading to elevated error rates that undermine practical utility.

The fundamental challenge addressed in this work stems from the inability of standard ASR architectures to explicitly differentiate between meaningful speech signals and irrelevant acoustic interference. This limitation manifests as increased word error rates and character error rates when processing audio with poor signal-to-noise characteristics.

This paper introduces an augmented architecture that extends the wav2vec2 model by incorporating a parallel noise detection pathway. Unlike conventional approaches that handle noise through preprocessing or post-processing stages, the proposed method integrates noise awareness directly into the feature learning process. This architectural modification enables the system to simultaneously optimize for both accurate transcription and reliable noise identification.

The primary contributions of this research include:

\begin{itemize}
\item Development of an integrated dual-head architecture combining transcription and noise classification capabilities
\item Systematic evaluation of multiple architectural configurations through controlled experiments
\item Demonstration of improved robustness metrics while maintaining transcription performance
\item Analysis of training strategies for joint optimization of multiple objectives
\end{itemize}

\section{RELATED WORK}

\subsection{Noise Robustness in Speech Recognition}

Addressing acoustic interference in speech recognition has been a persistent research focus across both neural network and traditional statistical approaches. Early methods emphasized front-end signal processing techniques such as spectral subtraction to enhance signal quality before recognition. Contemporary neural approaches have shifted toward end-to-end systems that implicitly learn noise-robust representations. Recent work has explored various strategies for improving robustness, including domain adaptation techniques, multi-condition training paradigms, and architectural modifications. Data augmentation using simulated room acoustics and diverse noise conditions has proven particularly effective for building generalized models. The availability of large-scale datasets containing realistic acoustic environments has enabled more comprehensive robustness evaluation.

\subsection{Self-Supervised Speech Models}

The wav2vec2 framework represents a significant advancement in speech representation learning through self-supervised pretraining on unlabeled audio data. By learning representations that capture acoustic and linguistic structure without explicit supervision, these models achieve strong performance with limited labeled data for downstream tasks.

Cross-lingual extensions of wav2vec2 have demonstrated the potential for multilingual speech recognition using shared architectures. However, performance in acoustically challenging conditions remains an active area of investigation, particularly for low-resource languages and noisy environments.

\subsection{Distinguishing Features of This Work}

While previous research has addressed noise robustness through various means, the approach presented here differs in several key aspects. Rather than treating noise handling as a preprocessing concern or relying solely on augmented training data, this work integrates noise awareness as an explicit architectural component. The dual-head design enables simultaneous optimization of transcription accuracy and noise detection, creating a system that inherently understands when input audio contains meaningful speech versus when it should abstain from generating transcriptions.

\section{PROPOSED METHODOLOGY}

\subsection{Dataset Construction}

The experimental framework utilizes multiple publicly available corpora to ensure comprehensive coverage of speech and noise conditions.

\subsubsection{Speech Audio Collections}

For English language speech content, three established datasets were selected: LibriSpeech provides read audiobook recordings totaling 66 hours for training, Common Voice supplies diverse speaker recordings for validation spanning 3 hours, and FLEURS offers multilingual speech data used for testing purposes with 1 hour of English content.

\subsubsection{Environmental Audio Collections}

The noise component incorporates three distinct sources: MUSAN provides music and ambient sounds, UrbanSound8K contributes urban environmental recordings, and office environment recordings capture workplace acoustic conditions. These sources ensure representation of diverse interference types including music, crowd noise, mechanical sounds, and general environmental disturbances.

\subsubsection{Data Preprocessing}

All audio underwent standardized preprocessing including resampling to 16 kHz, removal of punctuation marks, conversion to lowercase characters, and replacement of whitespace with pipe symbols. The training set composition includes 5 percent noise samples mixed with speech audio, while evaluation sets maintain equal proportions of speech and noise content to rigorously test classification capabilities.

Table~\ref{tab:datasets} summarizes the dataset specifications used in this research.

\begin{table}[htbp]
\centering
\caption{Dataset Specifications for Training and Evaluation}
\label{tab:datasets}
\begin{tabular}{@{}lllcc@{}}
\toprule
\textbf{Set} & \textbf{Speech Source} & \textbf{Noise Source} & \textbf{Speech (hrs)} & \textbf{Noise (hrs)} \\
\midrule
Train & LibriSpeech & MUSAN & 66.0 & 3.5 \\
Validation & Common Voice & MUSAN & 3.0 & 3.0 \\
Test & FLEURS & Office+Urban & 1.0 & 1.0 \\
\bottomrule
\end{tabular}
\end{table}

\subsection{Architectural Design}

The proposed architecture builds upon established self-supervised speech models with targeted modifications to enable noise awareness. Figures~\ref{fig:baseline} through~\ref{fig:fusion} illustrate the progressive architectural enhancements.

\subsubsection{Base Architecture Selection}

\begin{figure}[H]
\centering
\includegraphics[width=0.9\textwidth]{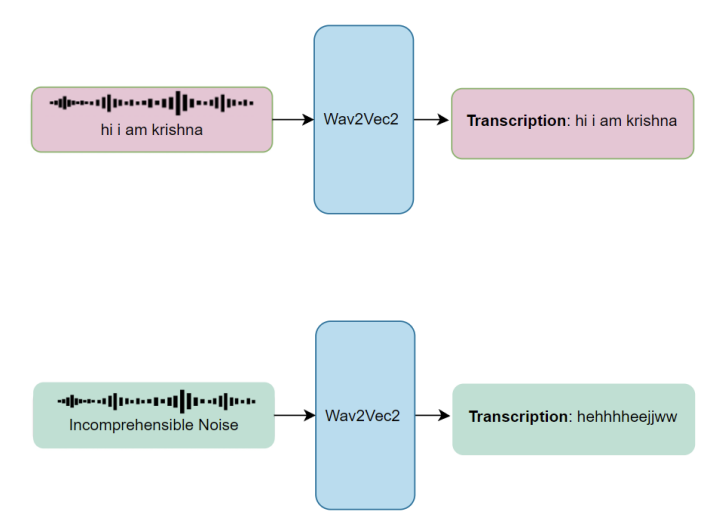}
\caption{Standard wav2vec2 architecture without noise handling capabilities.}
\label{fig:baseline}
\end{figure}

The foundation for this work is the wav2vec2-XLSR-53 model, which provides multilingual speech representations through cross-lingual pretraining. This base model was previously fine-tuned for English speech recognition, providing a strong starting point for further adaptation. The baseline configuration shown in Figure~\ref{fig:baseline} represents the standard approach without explicit noise handling.

\subsubsection{Noise Detection Integration}

\begin{figure}[H]
\centering
\includegraphics[width=0.9\textwidth]{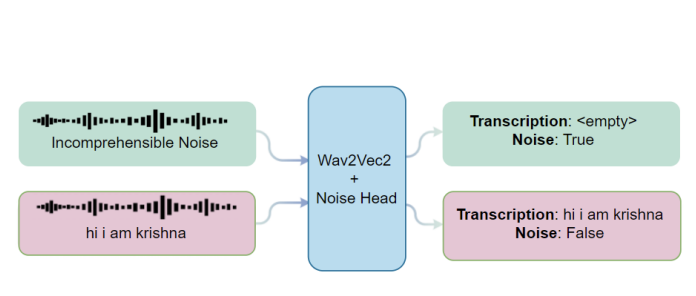}
\caption{Comparison showing enhanced model's noise classification capability versus conventional approach.}
\label{fig:comparison}
\end{figure}

\begin{figure}[H]
\centering
\includegraphics[width=\textwidth]{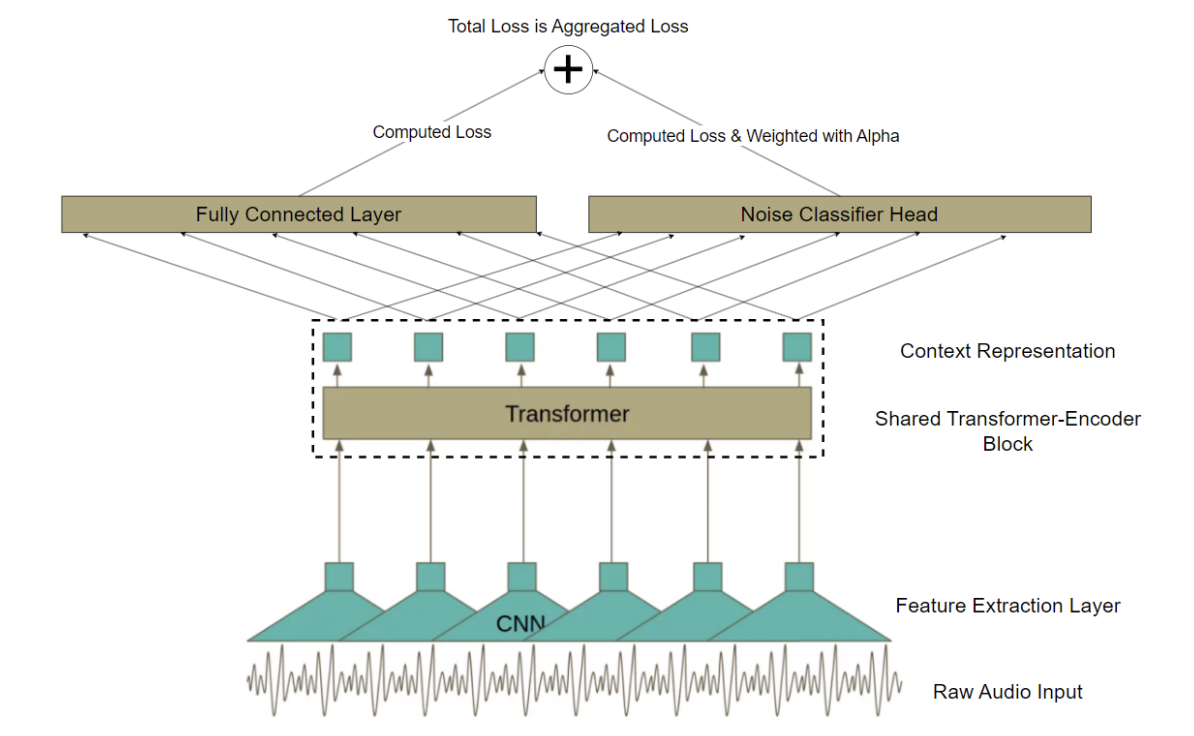}
\caption{Detailed architecture of wav2vec2 with integrated noise classification head.}
\label{fig:detailed}
\end{figure}

The core innovation involves adding a parallel classification pathway to the existing transcription decoder, as illustrated in Figures~\ref{fig:comparison} and~\ref{fig:detailed}. This noise detection head consists of a linear transformation layer followed by softmax activation, producing probability distributions over noise versus speech categories. The architectural modification enables the model to learn representations useful for both transcription and noise discrimination simultaneously. Unlike the baseline model, this enhanced architecture can explicitly identify when input audio contains non-speech content.

\subsubsection{Loss Function Design}

\begin{figure}[H]
\centering
\includegraphics[width=\textwidth]{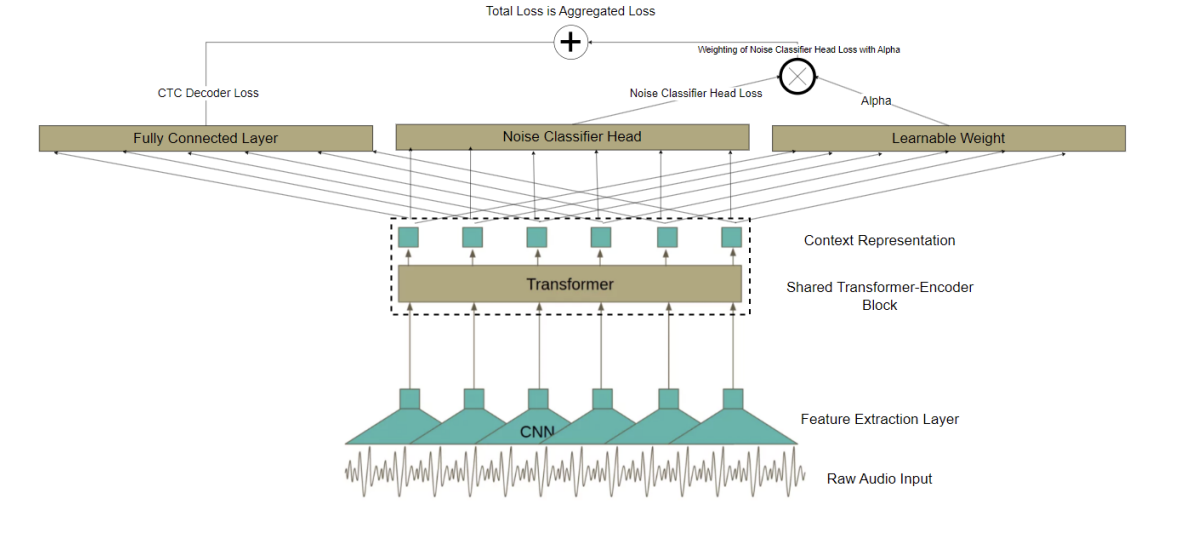}
\caption{Architecture incorporating adaptive loss weighting through trainable alpha parameter.}
\label{fig:adaptive}
\end{figure}

Training optimization combines two objective functions: connectionist temporal classification loss for transcription and cross-entropy loss for noise classification. The total training objective is computed as a weighted combination of these losses, where the relative weighting can be fixed or learned during training. Figure~\ref{fig:adaptive} demonstrates the architecture with trainable loss weighting parameter, enabling dynamic balance between transcription and classification objectives.

\subsubsection{Feature Fusion Strategies}

\begin{figure}[H]
\centering
\includegraphics[width=\textwidth]{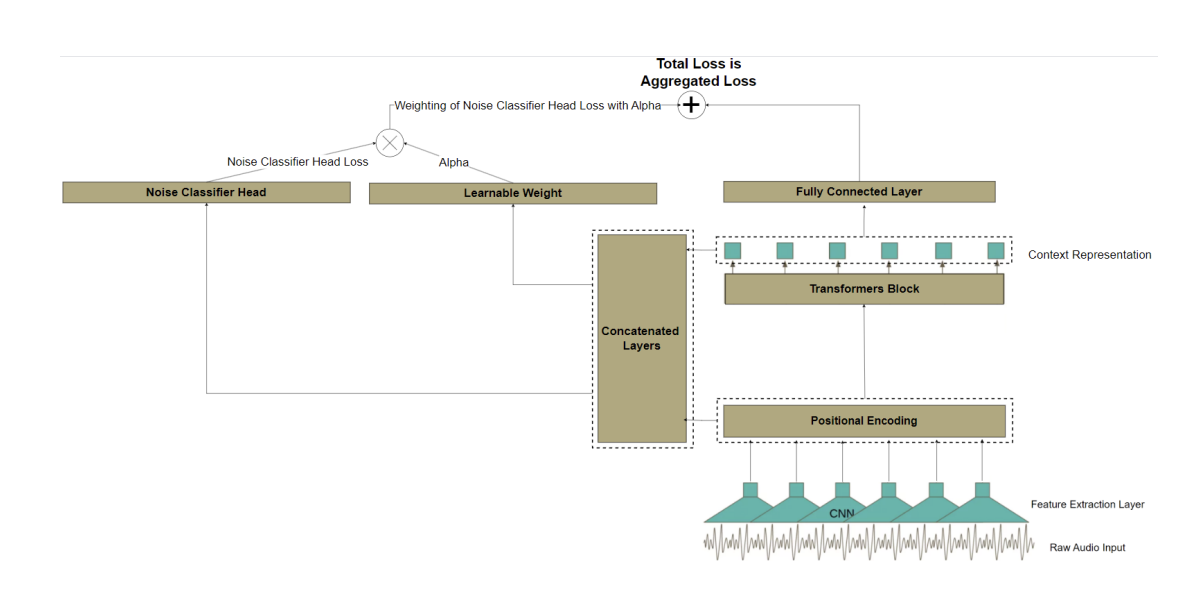}
\caption{Advanced feature fusion architecture concatenating CNN positional encodings with transformer context representations.}
\label{fig:fusion}
\end{figure}

Beyond basic dual-head architecture, experiments explored alternative feature combination approaches. One configuration incorporates positional encoding from the convolutional feature extractor combined with contextual representations from transformer layers, creating richer feature representations for both decoding pathways. Figure~\ref{fig:fusion} illustrates this advanced architecture where features from multiple processing stages are concatenated before classification.

\section{EXPERIMENTAL FRAMEWORK}

\subsection{Evaluation Data}

Two distinct test sets enable comprehensive performance assessment:

The speech test set consists entirely of FLEURS English recordings spanning one hour, providing ground truth for transcription accuracy measurement through word error rate and character error rate metrics.

The noise test set comprises one hour of purely non-speech audio including faint background sounds, incomprehensible vocalizations, and urban environmental recordings, enabling measurement of noise detection accuracy.

\subsection{Performance Metrics}

Three primary metrics quantify system performance:

\begin{itemize}
\item \textbf{Word Error Rate:} computed as the edit distance between predicted and reference transcriptions normalized by reference length
\item \textbf{Character Error Rate:} similar to WER but calculated at character level
\item \textbf{Noise Classification Accuracy:} percentage of correctly identified noise versus speech samples
\end{itemize}

For baseline models lacking explicit noise classification heads, noise detection accuracy was inferred from transcription output, considering empty transcriptions as noise detection.

\subsection{Reference Baseline}

The baseline system consists of the wav2vec2-XLSR-53 model fine-tuned on LibriSpeech training data and validated on Common Voice. This configuration represents standard practice for adapting pretrained models to speech recognition tasks without explicit noise handling mechanisms.

\subsection{Experimental Configurations}

Four distinct experiments systematically evaluate architectural variants:

\textbf{Configuration A:} Standard fine-tuning of the base model on mixed speech-noise training data establishes baseline noise classification performance for conventional architectures.

\textbf{Configuration B:} Addition of noise classification head with fixed loss weighting (alpha = 0.01) tests the basic dual-head architecture.

\textbf{Configuration C:} Extension of Configuration B making the loss weighting parameter trainable during optimization, enabling automatic balance between objectives.

\textbf{Configuration D:} Further extension incorporating feature concatenation from convolutional and transformer layers to provide richer representations.

\subsection{Training Details}

All experiments employed consistent training hyperparameters: batch size of 8 samples, learning rate of 1e-5 with Adam optimizer, and 30 training epochs. Training was conducted on NVIDIA A100 80GB GPUs. The training and validation datasets remained constant across all configurations to ensure fair comparison.

\section{RESULTS AND ANALYSIS}

Table~\ref{tab:results} presents comprehensive results across all experimental configurations and evaluation metrics.

\begin{table}[htbp]
\centering
\caption{Performance Comparison Across Configurations}
\label{tab:results}
\begin{tabular}{@{}lccc@{}}
\toprule
\textbf{Configuration} & \textbf{Noise Acc (\%)} & \textbf{WER (\%)} & \textbf{CER (\%)} \\
\midrule
Baseline & 6.0 & 11.85 & 4.40 \\
Configuration A & 99.3 & 14.15 & 5.20 \\
Configuration B & 99.3 & 11.43 & 4.37 \\
Configuration C & 99.8 & 11.76 & 4.44 \\
Configuration D & 98.3 & 11.88 & 4.46 \\
\bottomrule
\end{tabular}
\end{table}

\subsection{Noise Classification Performance}

The baseline system demonstrates minimal noise detection capability at 6 percent accuracy, confirming that standard architectures lack inherent mechanisms for distinguishing noise from speech. Configuration A achieves dramatic improvement to 99.3 percent accuracy through training on mixed data, validating that exposure to noise examples enables discrimination. However, this comes at the cost of degraded transcription performance.

Configurations B through D maintain high noise classification accuracy above 98 percent while addressing the transcription degradation observed in Configuration A. This demonstrates that explicit architectural support for noise classification, rather than implicit learning alone, enables joint optimization of both objectives.

\subsection{Transcription Quality}

Configuration A exhibits the poorest transcription metrics with WER of 14.15 percent and CER of 5.20 percent, representing degradation from the baseline. This illustrates the challenge of learning from mixed data without architectural modifications.

Configuration B recovers transcription performance, achieving WER of 11.43 percent and CER of 4.37 percent, actually surpassing baseline performance. This indicates that proper architectural design enables the model to leverage noise training data constructively rather than destructively.

Configuration C with trainable loss weighting shows slight WER increase to 11.76 percent but maintains competitive CER at 4.44 percent while achieving the highest noise accuracy at 99.8 percent. This suggests the optimization process finds different balances between objectives when loss weighting is flexible.

Configuration D shows minimal difference from Configuration C, suggesting that feature concatenation does not provide substantial additional benefit in this setting.

\subsection{Key Findings}

Several important observations emerge from these results:

First, incorporating noise training data without architectural support degrades transcription performance, as evidenced by Configuration A results. Second, adding an explicit noise classification head enables effective joint optimization, as demonstrated by Configuration B maintaining transcription quality while achieving high noise accuracy. Third, flexible loss balancing through trainable weighting provides marginal additional improvements in noise classification with minimal transcription impact.

Overall, the results validate the proposed architectural approach for building noise-aware speech recognition systems that maintain strong transcription performance while gaining robust noise detection capabilities.

\section{CONCLUSION AND FUTURE DIRECTIONS}

This research demonstrates that integrating noise classification directly into ASR architecture yields systems with enhanced robustness and awareness of input signal quality. The proposed dual-head design enables simultaneous optimization of transcription and noise detection objectives, avoiding the performance degradation that occurs when training on mixed data with conventional architectures.

Experimental results confirm that appropriate architectural modifications enable models to benefit from noise training examples without compromising transcription accuracy. The best configuration achieves near-perfect noise classification while maintaining word error rates competitive with or superior to baseline systems.

These findings suggest several promising directions for future investigation. Extension to multilingual settings would validate whether similar benefits accrue across diverse languages. Exploration of alternative feature fusion strategies beyond simple concatenation might yield additional performance gains. Investigation of adaptive inference mechanisms that adjust behavior based on detected noise levels could further enhance practical utility.

The approach presented here provides a foundation for building more robust speech recognition systems that explicitly reason about input signal quality, enabling more reliable deployment in challenging real-world acoustic environments.

\section*{REFERENCES}

\begin{enumerate}
\item Testing with Kolena, ``WER, CER, and MER,'' Technical Documentation, 2024.

\item A. Conneau, A. Baevski, R. Collobert, A. Mohamed, and M. Auli, ``Unsupervised Cross-Lingual Representation Learning for Speech Recognition,'' arXiv preprint arXiv:2006.13979, 2020.

\item A. Baevski, H. Zhou, A. Mohamed, and M. Auli, ``wav2vec 2.0: A Framework for Self-Supervised Learning of Speech Representations,'' \textit{Advances in Neural Information Processing Systems}, vol. 33, pp. 12449--12460, 2020.

\item J. Li, L. Deng, Y. Gong, and R. Haeb-Umbach, ``An Overview of Noise-Robust Automatic Speech Recognition,'' \textit{IEEE/ACM Transactions on Audio, Speech, and Language Processing}, vol. 22, no. 4, pp. 745--777, 2014.

\item V. Panayotov, G. Chen, D. Povey, and S. Khudanpur, ``Librispeech: An ASR Corpus Based on Public Domain Audio Books,'' in \textit{IEEE International Conference on Acoustics, Speech and Signal Processing (ICASSP)}, pp. 5206--5210, 2015.

\item R. Ardila et al., ``Common Voice: A Massively-Multilingual Speech Corpus,'' in \textit{Proceedings of the 12th Conference on Language Resources and Evaluation (LREC 2020)}, pp. 4211--4215, 2020.

\item A. Conneau et al., ``FLEURS: Few-shot Learning Evaluation of Universal Representations of Speech,'' arXiv preprint arXiv:2205.12446, 2022.

\item D. Snyder, G. Chen, and D. Povey, ``MUSAN: A Music, Speech, and Noise Corpus,'' arXiv preprint arXiv:1510.08484, 2015.

\item J. Salamon, C. Jacoby, and J. P. Bello, ``A Dataset and Taxonomy for Urban Sound Research,'' in \textit{22nd ACM International Conference on Multimedia}, pp. 1041--1044, 2014.

\item A. Graves, S. Fernandez, F. Gomez, and J. Schmidhuber, ``Connectionist Temporal Classification: Labelling Unsegmented Sequence Data with Recurrent Neural Networks,'' in \textit{Proceedings of the 23rd International Conference on Machine Learning}, pp. 369--376, 2006.
\end{enumerate}

\end{document}